\documentclass[conference]{IEEEtran}
\IEEEoverridecommandlockouts  
\usepackage[letterpaper, left=0.68in, right=0.68in, top=0.75in, bottom=1.1in]{geometry}

\usepackage{amsmath,amsfonts}

\usepackage{algorithmic}
\usepackage{array}
\usepackage{times}  

\usepackage{amsmath,array,float}
\usepackage{xcolor}

\newcommand{\gmkn}{g_{mk}^{(n)}}
\newcommand{\hmkn}{h_{mk}^{(n)}}

\usepackage{balance}

\usepackage{booktabs,       
            makecell,       
            tabularx}       
\usepackage{multirow}
\usepackage{multicol}
\usepackage[export]{adjustbox}
\usepackage{balance} 
\usepackage{tikz}
\usetikzlibrary{shapes.geometric}
\usetikzlibrary{arrows}
\usetikzlibrary{fit}
\usetikzlibrary{calc}
\tikzstyle{startstop} = [rectangle, rounded corners, minimum width=3cm, minimum height=1cm,text centered, text width=3cm, draw=black]
\tikzstyle{io} = [trapezium, trapezium left angle=70, trapezium right angle=110, minimum width=3cm, minimum height=1cm, text centered, text width=3cm, draw=black]
\tikzstyle{process} = [rectangle, minimum width=3cm, minimum height=1cm, text centered, text width=3cm, draw=black]
\tikzstyle{decision} = [diamond, aspect=3, align=center, inner sep=-1ex, minimum width=2cm, minimum height=1cm, text centered, text width=5cm, draw=black]
\tikzstyle{arrow} = [thick,->,>=stealth]

\def\BibTeX{{\rm B\kern-.05em{\sc i\kern-.025em b}\kern-.08em
    T\kern-.1667em\lower.7ex\hbox{E}\kern-.125emX}}
\begin{document}
%
\title{Federated Learning for Secure and Efficient Device Activity Detection in mMTC Networks}

\author{\IEEEauthorblockN{Ali Elkeshawy\IEEEauthorrefmark{1}, Ibrahim Al Ghosh\IEEEauthorrefmark{1}, Ha\"{\i}fa Far\`es\IEEEauthorrefmark{1}, Amor Nafkha\IEEEauthorrefmark{1} \\}
\IEEEauthorblockA{\IEEEauthorrefmark{1} IETR - UMR CNRS 6164, CentraleSup\' elec, avenue de la Boulaie - CS 47601 35576 \\ CESSON-SEVIGNE Cedex, France\\
Email: \{ali-fekry-ali-hassan.elkeshawy, ibrahim.al-ghosh, haifa.fares, amor.nafkha\}@centralesupelec.fr}}
\maketitle
\begin{abstract}
Grant-free random access in massive machine-type communications  enables low-latency connectivity with minimal signaling. However, sporadic device activation requires efficient device activity detection. We propose a federated learning-based device activity detection approach, leveraging distributed training to enhance security and privacy while maintaining low computational complexity. Compared to existing methods, our solution achieves competitive detection performance, addressing scalability and security challenges in mMTC networks.
\end{abstract}

\IEEEpeerreviewmaketitle

\section{Introduction}
\noindent The rapid growth of massive machine-type communications (mMTC) in 5G and beyond networks has introduced significant challenges, including sporadic device activity, low-latency requirements, and the need for energy-efficient solutions. Cell-free massive MIMO (CF-mMIMO) has emerged as a promising architecture to address these challenges, offering improved coverage and scalability by deploying distributed access points (APs) that collaboratively serve devices. In this context, grant-free random access (GF-RA) has become a key enabler, allowing devices to transmit data without prior scheduling, thereby reducing signaling overhead and improving network efficiency. However, efficient device activity detection (AD) remains a critical component for accurate data decoding in mMTC systems. Traditional approaches, such as compressive sensing and covariance-based methods, often struggle with scalability and computational complexity, especially in large-scale deployments.

\begin{figure}
  \centering
  \includegraphics[width=\columnwidth,height=5cm]{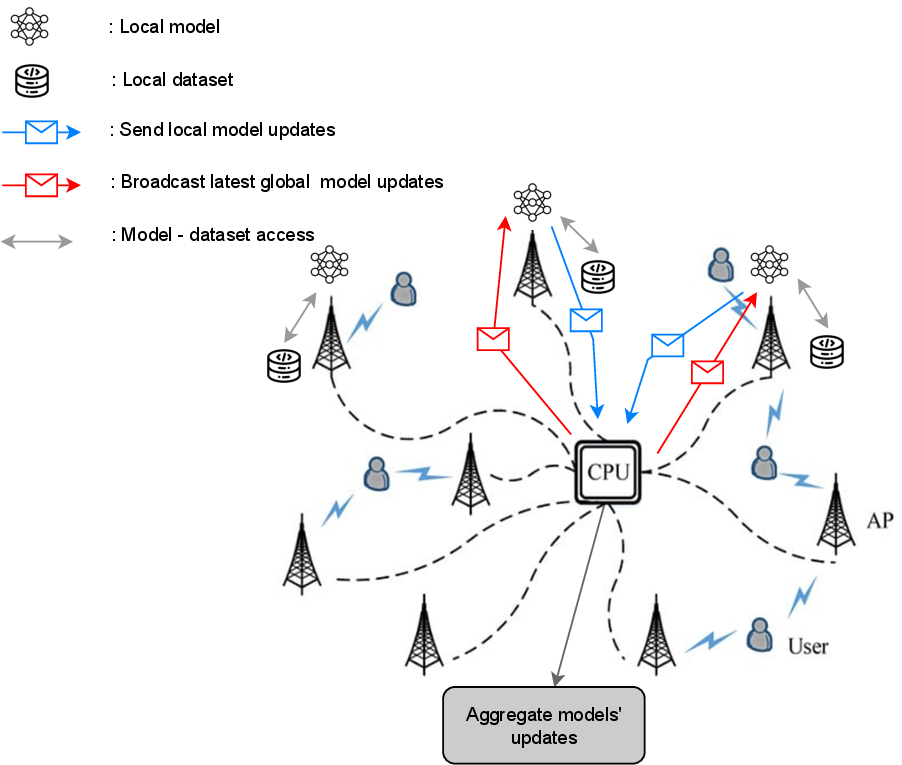}
  \vspace{0.5pt} 
  \caption{CF-mMIMO federated learning framework with secure aggregation updates.}
  \vspace{-10pt} 
  \label{fig1}
\end{figure}

 Federated learning (FL) has recently gained attention as a promising framework for distributed machine learning, particularly in scenarios where data privacy and scalability are paramount \cite{fedlearning}. In the context of CF-mMIMO, FL offers a decentralized approach to resolve the problem of AD, enabling APs to collaboratively train a global model without sharing raw data. Each AP uses its locally received signals to compute model updates, which are aggregated by a central server to refine the global model, see Fig.\ref{fig1}. This iterative process not only preserves data privacy but also enhances scalability and adaptability to dynamic network conditions.

\section{System Model}
\noindent We analyze a CF-mMIMO network with $M$ APs, each featuring $N$ antennas, serving $K$ single-antenna devices within a square area of $D \times D$ km\textsuperscript{2}, as described in \cite{Elkeshawy2025}. All APs are connected to a central processing unit (CPU). Due to the sporadic nature of traffic in mMTC, only a small subset of devices is active at any given time. The activity of the \( k \)-th device is represented by \( a_k \in \{0, 1\} \), where \( a_k = 1 \) indicates activity and \( a_k = 0 \) inactivity. The activation probability \( \epsilon \ll 1 \) follows a bernoulli distribution, resulting in a sparse activity vector \( \mathbf{a} = (a_1, a_2, \dots, a_K) \). The channel gain $\gmkn$ between the \( n \)-th antenna of the \( m \)-th AP and the \( k \)-th device is modeled as:
\begin{equation}
\label{eq:Channelfad}
\gmkn = \beta_{mk}^{1/2} \times \hmkn,
\end{equation}

where \( \beta_{mk} \) represents the large-scale fading coefficient, and $\hmkn$ captures small-scale fading.
Each active device transmits a non-orthogonal pilot sequence \( \mathbf{s}_k \in \mathbb{C}^{L \times 1} \), where \( L \) is the pilot sequence length.
The signal received at the \(m\)-th AP during the uplink phase, \(\mathbf{Y}_m \in \mathbb{C}^{L \times N}\), from \cite{Elkeshawy2025} is be represented as follows:
\begin{equation}
\label{eq:SignalrecN}
\mathbf{y}_{mn} = \sum_{k=1}^{K} a_k \rho_k^{1/2} g_{mk}^{(n)} s_k + w_{mn},
\end{equation}

where \( \rho_k \) is the transmit power of the \( k \)-th device, and \( \mathbf{w}_{mn} \sim \mathcal{CN}(0, \sigma^2 \mathbf{I}_L) \) is the additive white gaussian noise (AWGN) vector.
For the rest, we consider the signal received at the \(m\)-th AP during the uplink phase, \(\mathbf{Y}_m \in \mathbb{C}^{L \times N}\), from \cite{Elkeshawy2025}.
This model captures the key aspects of the CF-mMIMO uplink scenario, emphasizing the sporadic activity of devices and the distributed nature of the network.

\section{Federated Learning for Device Activity Detection}

\noindent The primary objective of the device AD model is to estimate the activity indicator vector \(\mathbf{a}\) from the received signal \(\mathbf{Y}_m\). To achieve this, we employ FL, a decentralized machine learning approach where multiple APs collaboratively train a local model while keeping their data distributed rather than storing it centrally. This is particularly beneficial in massive MIMO networks, where devices generate heterogeneous and non-iid  data due to varying channel conditions and traffic patterns.

In our setup, each AP trains an identical local model using its received data. Instead of transmitting raw data, each AP computes local model updates  parameters) based on its observations and periodically sends them to the CPU. The CPU aggregates these updates using weighted averaging, ensuring that APs with higher-quality data contribute more significantly to the global model. Once aggregated, the CPU performs a global optimization step before distributing the updated model back to the APs, allowing for continuous improvement across multiple training rounds.

The model architecture is inspired by \cite{Elkeshawy2025} and follows a single layer perceptron (SLP) structure, designed for efficiency and fast convergence in large-scale networks. The model consists of an input layer that represents the received signals \(\mathbf{Y}_m\), followed by a single hidden layer with \(V = 512\) artificial neurons. Unlike deeper architectures, it features only one hidden layer (\(Z = 1\)), making it computationally lightweight. The output layer generates probability values \(\mathbf{\tilde{a}}\) in the interval \([0,1]\), which are processed through a hard-thresholding step to obtain the estimated device activity. To improve decision reliability, a ponderation process is carried out at the CPU, selecting a cluster of the most reliable APs (of size \(T\)) for each device before making a final classification.

During training, both the APs and the CPU utilize the Adam optimizer, known for its adaptive learning rate and fast convergence. The binary cross-entropy loss function is minimized using the backpropagation technique, where gradients are propagated backward to refine the model parameters. This iterative learning process allows the model to improve device detection accuracy while ensuring robustness against noisy channel conditions and limited communication resources.

\section{Numerical Result}
\noindent The uplink mMTC system is modeled within a \(1 \text{ km}^2\) square region, comprising \( M = 20 \) APs, each equipped with \( N = 2 \) antennas. The network consists of \( K = 100 \) devices, each activating with a probability of \( \epsilon = 10\% \), the cluster size is set to \( T = 4 \). The non-orthogonal pilot sequences have a length of \( L = 40 \). The performance of device AD is evaluated using the receiver operating characteristic (ROC) curve, which represents the relationship between the true positive rate (TPR) and the false positive rate (FPR).

\begin{figure}
  \centering
  \includegraphics[width=\columnwidth,height=5cm]{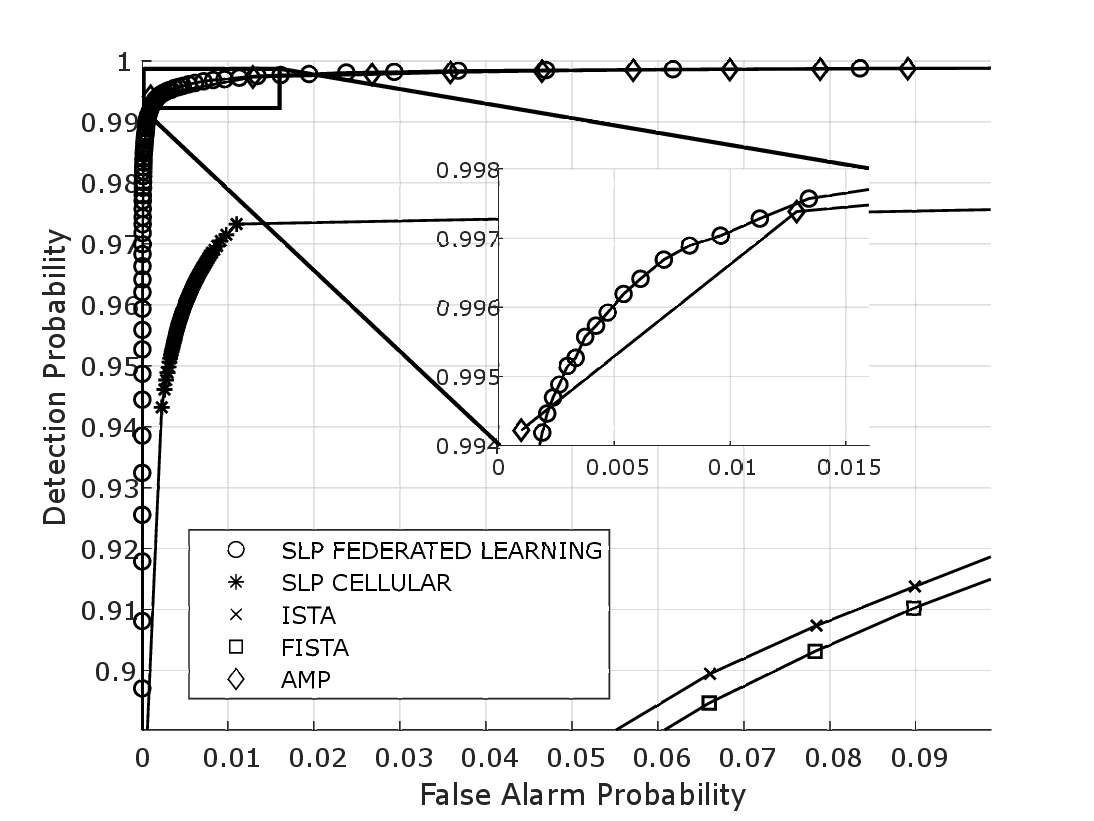}
  \caption{ROC curves: federated learning vs. literature benchmarks.}
  \vspace{-5pt} 
  \label{fig2}
\end{figure}


Fig.~\ref{fig2} presents a comparative analysis of the proposed FL-based approach against well-established algorithms, including approximate message passing (AMP), the iterative soft thresholding algorithm (ISTA), and its accelerated variant, fast-ISTA (FISTA) \cite{AMPtutorial}. These methods are widely used for sparse problem recovery, yet the results demonstrate that FL consistently outperforms them in terms of detection probability across various false alarm rates. Furthermore, the results highlight the advantages of CF-mMIMO over the colocated (cellular) scenario, showcasing its ability to enhance detection performance due to its distributed nature. In addition to performance gains, AMP is computationally expensive, requiring approximately six times more multiply-accumulate (MAC) operations compared to the proposed FL-based approach. Beyond these improvements, FL also introduces an additional layer of security and privacy by ensuring that APs transmit only model parameters to the CPU rather than raw received signals. This decentralized training approach preserves data confidentiality while simultaneously reducing communication overhead, making it a more robust and efficient solution for large-scale mMTC networks.

\section{Conclusion}
\noindent This work proposed a FL-based device AD approach for CF-mMIMO, enhancing security, scalability, and detection accuracy. Results show superior performance over state-of-the-art sparse recovery methods and colocated architectures. By exchanging model parameters instead of raw signals, the framework ensures privacy while optimizing network efficiency.
\section*{Acknowledgment}
\noindent This work received funding from the French National Research Agency (ANR-22-CE25-0015) within the frame of the project POSEIDON.

\bibliography{Main}
\bibliographystyle{IEEEtran}
\end{document}